\documentstyle[preprint,aps]{revtex}

\begin{document}
\draft

\title{Three-body correlations in few-body nuclei}

\author{A.\ Arriaga}
\address{Centro de Fisica Nuclear da Universidade de Lisboa,
         Avenida Gama Pinto 2, 1699 Lisboa, Portugal}
\author{V.R.\ Pandharipande}
\address{Department of Physics, University of Illinois,
         Urbana, IL 61801, USA}
\author{R.B.\ Wiringa}
\address{Physics Division, Argonne National Laboratory,
         Argonne, IL 60439, USA}

\date{\today}
\maketitle

\begin{abstract}
A detailed comparison of Faddeev and variational wave functions for $^3$H,
calculated with realistic nuclear forces, has been made to study the form
of three-body correlations in few-body nuclei.
Three new three-body correlations for use in variational wave functions
have been identified, which substantially reduce the difference with the
Faddeev wave function.
The difference between the variational upper bound and the Faddeev binding
energy is reduced by half, to typically $<2\%$.
These three-body correlations also produce a significant lowering of the
variational binding energy for $^4$He and larger nuclei.
\end{abstract}

\narrowtext

\section{INTRODUCTION}

Quantum Monte Carlo methods, adopted for nuclei in the last decade, have
proven to be very useful in obtaining exact solutions of the nuclear
Schr\"odinger equation for the ground and low-energy excited states of up to
six nucleons~\cite{PPCW95}, and it is likely that many states of seven and
eight nucleons will be exactly calculated with realistic nuclear forces in
the near future. A variational approximation, $|\Psi_v \rangle$, to the
lowest-energy state of the desired spin, parity, and isospin, $(J^{\pi},T)$,
is first obtained by a variational Monte Carlo (VMC) calculation~\cite{W91}.
The exact eigenstate, $|\Psi_0 \rangle$, belonging to the lowest-energy
eigenvalue, $E_0$, for that $(J^{\pi},T)$ is then projected out using
\begin{equation}
|\Psi_0\rangle = \displaystyle\lim_{\tau\to\infty}
                      exp[-(H-E_0)\tau] |\Psi_v\rangle
\end{equation}
with the Green's function Monte Carlo (GFMC) method~\cite{C88}.
The computation yields a population of configurations ${\bf R}_i$, where
${\bf R} = ({\bf r}_1,{\bf r}_2,\cdots,{\bf r}_A)$ specifies the positions
of all the nucleons and $i=1,N_c$ labels the configurations, distributed
with the probability
\begin{equation}
     P({\bf R},\tau) \propto |\Psi^{\dagger}_v({\bf R})
                     exp[-(H-E_0)\tau] \Psi_v({\bf R})| \ .
\end{equation}
Mixed estimates, defined as:
\begin{equation}
     [O(\tau)]_M = \frac{ \langle\Psi_v| O exp[-(H-E_0)\tau] |\Psi_v\rangle }
                        { \langle\Psi_v| exp[-(H-E_0)\tau] |\Psi_v\rangle } \ ,
\end{equation}
are then calculated from this population.
Obviously $[H(\tau)]_M$ converges to the eigenvalue $E_0$ in the limit
$\tau\to\infty$, however, the variance of $[H(\tau)]_M$ depends primarily
on that of the local variational energy:
\begin{equation}
     E_v({\bf R}) =\frac{ \Psi^{\dagger}_v({\bf R}) H \Psi_v({\bf R})}
                         { \Psi^{\dagger}_v({\bf R})   \Psi_v({\bf R}) } \ .
\end{equation}
Expectation values of operators other than H are generally calculated only
up to first order in the difference $(|\Psi_0 \rangle - |\Psi_v \rangle)$ from
\begin{equation}
     \langle O \rangle = 2 \displaystyle\lim_{\tau\to\infty} [O(\tau)]_M
                       - \frac{ \langle\Psi_v| O |\Psi_v\rangle }
                              { \langle\Psi_v|\Psi_v\rangle } \ .
\end{equation}
It is clear that accurate variational wave functions are needed for the
success of this approach.
The variance of $E_v({\bf R})$ decreases and the terms neglected in eq.(5)
become smaller as $|\Psi_v \rangle \to |\Psi_0 \rangle$.

State-of-the-art variational wave functions~\cite{W91} of few-body nuclei
have the form:
\begin{equation}
     |\Psi_v\rangle = [ 1 + \sum_{i<j}U^{LS}_{ij} + \sum_{i<j<k}U^{TNI}_{ijk} ]
                      |\Psi_p\rangle \
\end{equation}
with the pair wave function $|\Psi_p\rangle$ given by:
\begin{equation}
     |\Psi_p\rangle = [ {\it S}\prod_{i<j}(1+U_{ij}) ] |\Psi_{J}\rangle \ .
\end{equation}
The ${\displaystyle S\prod_{i<j}}$ represents a symmetrized product, and the
Jastrow wave function, $|\Psi_{J}\rangle$, is given by:
\begin{equation}
     |\Psi_J\rangle = \left[ \prod_{i<j}f_c(r_{ij}) \right] |\Phi\rangle \ .
\end{equation}
Here $|\Phi\rangle$ is an antisymmetric product of single-particle wave
functions with the desired $(J^{\pi},T)$, the $f_c(r_{ij})$ is a
two-body central correlation, and the operator $U_{ij}$ is defined as:
\begin{equation}
     U_{ij} = \sum_{p=2,6} \left[ \prod_{k\not=i,j}f^p_{ijk}({\bf r}_{ik}
              ,{\bf r}_{jk}) \right] u_p(r_{ij}) O^p_{ij} \ ,
\end{equation}
\begin{equation}
     O^{p=1,6}_{ij} = [ 1, \sigma_i\cdot\sigma_j, S_{ij} ]\otimes
                      [ 1, \tau_i\cdot\tau_j ] \ .
\end{equation}
The $f_c(r_{ij})$ and $U_{ij}$ correlations are generated by the static parts
of the two-nucleon interaction.
The spin-orbit correlations are defined as:
\begin{equation}
     U^{LS}_{ij} = [ u_{\ell s}(r_{ij}) + u_{\ell s\tau}(r_{ij})
                   \tau_i\cdot\tau_j ] ( {\bf L}\cdot{\bf S} )_{ij} \ ,
\end{equation}
and the eight radial functions $f_c(r_{ij})$, $u_{p=2,6}(r_{ij})$,
$u_{\ell s}(r_{ij})$, and $u_{\ell s\tau}(r_{ij})$ are obtained from
approximate
two-body Euler-Lagrange equations with variational parameters~\cite{W91}.
The factor:
\begin{equation}
     f^p_{ijk} = 1 - t_1\left(\frac{r_{ij}}{R_{ijk}}\right)^{t_2}
                        exp[-t_3R_{ijk}] \ ,
\label{eq:fpijk}
\end{equation}
with
\begin{equation}
     R_{ijk} = r_{ij} + r_{jk} + r_{ki} \ ,
\end{equation}
suppresses spin-isospin correlations between nucleons $i$ and $j$ when a third
nucleon $k$ comes close to either $i$ or $j$, and $t_{1-3}$ are variational
parameters. Finally,
\begin{equation}
     U^{TNI}_{ijk} = \epsilon V_{ijk}(\overline{ r}_{ij},
                   \overline{r}_{jk}, \overline{ r}_{ki}) \ ,
\end{equation}
with $\overline{r}=br$, represent correlations induced by the three-nucleon
interaction $V_{ijk}(r_{ij},r_{jk},r_{ki})$.
Its form is suggested by perturbation theory, and $\epsilon$ and $b$ are
variational parameters.

The exact ground-state energy of $^3$H for the Argonne v$_{14}$
two-nucleon~\cite{WSA84} and Urbana model VIII three-nucleon~\cite{W91}
interactions, calculated using a Faddeev wave function~\cite{CPFG85} and Monte
Carlo integration, is $-8.49(1)$, while the energy for $^4$He, calculated
by the GFMC~\cite{C91} method, is $-28.3(2)$ MeV, where the numbers in
parentheses denote the statistical error.
VMC calculations~\cite{W91} with the wave function of eq.(6) give instead
the energies $-8.21(2)$ and $-27.23(6)$ MeV that are too high by $\sim 3.5\%$.
This difference, reflecting the deficiencies in $|\Psi_v \rangle$, could be
because the two-body functions $f_c$, $u_p$, $u_{\ell s}$, and $u_{\ell s\tau}$
are not optimally determined by the approximate equations used to construct
them, or the true $|\Psi_0 \rangle$ contains many-body correlations in addition
to $f^p_{ijk}$ and $U^{TNI}_{ijk}$.
Note that a $|\Psi_v \rangle$ containing two- and three-body correlations can
in principle reproduce the exact $|\Psi_0 \rangle$ of $^3$H.

We carried out extensive comparisons between the 34-channel Faddeev
$|\Psi_F\rangle$ of the Los Alamos-Iowa group~\cite{CPFG85}, one of the best
available approximations to the exact $|\Psi_0 \rangle$, and the
$|\Psi_v\rangle$ for $^3$H, to test the accuracy of the two-body functions
in $|\Psi_v\rangle$ and study the three-body correlations in $|\Psi_F\rangle$,
and therefore in $|\Psi_0 \rangle$.
These comparisons are discussed in Sec. II.
Slight improvements in the two-body functions did not significantly reduce the
difference $(|\Psi_F\rangle - |\Psi_v\rangle)$.
However, a few three-body correlation operators can reduce this difference
substantially.
VMC calculations, for the binding energy of the three and four-body systems,
with wave functions including the new correlations are presented in Sec. III.
The error in these is approximately half of that in the best previous
variational calculations~\cite{W91}.

\section{COMPARISON OF THE FADDEEV AND VARIATIONAL WAVE FUNCTIONS
AND THE NEW THREE-BODY CORRELATIONS}

The new variational wave function we adopt is given by:
\begin{equation}
     |\Psi_v\rangle = [ 1 + \sum_{i<j<k}U_{ijk} + \sum_{i<j}U^{LS}_{ij}
                          + \sum_{i<j<k}U^{TNI}_{ijk} ]
                      [ \prod_{i<j<k}f^c_{ijk} ] |\Psi_p\rangle \ ,
\end{equation}
and our purpose is to study the new three-body correlations, $f^c_{ijk}$ and
$U_{ijk}$, and also to improve the $f^p_{ijk}$ correlations in
$|\Psi_p\rangle$.
We consider the triton wave function expressed in the spin-isospin basis
\begin{equation}
     |\alpha \beta\rangle = |s_{12} s m\rangle
                            |t_{12} \frac{1}{2} \overline{\frac{1}{2}}\rangle \
 ,
\end{equation}
where $s_{12}$ and $t_{12}$ are respectively the total spin and isospin of
the pair of particles (12), $s$ is the total spin of the three nucleons, and
$m$ is its projection.
Allowed values for $s$ are $\frac{1}{2}$ (for $s_{12}=0,1$) and $\frac{3}{2}$
(for $s_{12}=1$).
The total isospin of the system is $\frac{1}{2}$, its projection is
$\overline{\frac{1}{2}} \equiv -\frac{1}{2}$, and $t_{12}$ can have the
values $0$ or $1$.
Consequently there are 16 components in the wave function, corresponding to the
8 spin states $|s_{12}sm\rangle$ labeled by $\alpha=1,8$ and the 2 isospin
states $|t_{12} \frac{1}{2} \overline{\frac{1}{2}}\rangle$ labeled by
$\beta=1,2$.
These are defined as:
\begin{eqnarray}
 &&  |1 \beta \rangle = |0 \frac{1}{2} \frac{1}{2}\rangle  |\beta\rangle \ ,
 \nonumber\\
 &&  |2 \beta \rangle = |0 \frac{1}{2} \overline{\frac{1}{2}}\rangle
 |\beta\rangle \ , \nonumber\\
 &&  |3 \beta \rangle = |1 \frac{1}{2} \frac{1}{2}\rangle  |\beta\rangle \ ,
 \nonumber\\
 &&  |4 \beta \rangle = |1 \frac{1}{2} \overline{\frac{1}{2}}\rangle
 |\beta\rangle \ , \nonumber\\
 &&  |5 \beta \rangle = |1 \frac{3}{2} \frac{3}{2}\rangle  |\beta\rangle \ ,
 \nonumber\\
 &&  |6 \beta \rangle = |1 \frac{3}{2} \frac{1}{2}\rangle  |\beta\rangle \ ,
 \nonumber\\
 &&  |7 \beta \rangle = |1 \frac{3}{2} \overline{\frac{1}{2}}\rangle
 |\beta\rangle \ , \nonumber\\
 &&  |8 \beta \rangle = |1 \frac{3}{2} \overline{\frac{3}{2}}\rangle
 |\beta\rangle \ ,
\end{eqnarray}
and
\begin{eqnarray}
 &&  |\alpha 1\rangle = |\alpha\rangle |0 \frac{1}{2}
 \overline{\frac{1}{2}}\rangle \ , \nonumber \\
 &&  |\alpha 2\rangle = |\alpha\rangle |1 \frac{1}{2}
 \overline{\frac{1}{2}}\rangle \ .
\end{eqnarray}
The triton wave function at any configuration ${\bf R}$ can then be written as
\begin{equation}
     \Psi({\bf R}) = \sum_{\alpha,\beta} \psi_{\alpha \beta}
                     ( {\bf R})|\alpha \beta\rangle
\label{eq:psive}
\end{equation}
The main components of the wave function are $ \psi_{31}$ and $ \psi_{12}$,
which are the only nonzero components in the Jastrow wave function.
Other components are generated mostly by tensor and three-body correlations.

In order to determine the $f^c_{ijk}$ and $U_{ijk}$ we compared
$|\Psi_v\rangle$ with $|\Psi_F\rangle$, and studied the deviations of the
former relative to the latter.
Using the decomposition of eq.(\ref{eq:psive}) for both
wave functions, we defined a number of ``deviation'' quantities.
For each spatial configuration, ${\bf R}_k$, the deviation
${\cal E}_{{\bf R}_k}$ is given by
\begin{equation}
{\cal E}_{{\bf R}_k} = \sum_{\alpha\beta} | \psi_{v,\alpha\beta}({\bf R}_k)
                         - \psi_{F,\alpha\beta} ({\bf R}_k) |^2 \ .
\label{eq:er}
\end{equation}
Separately, we define a deviation ${\cal E}_{\alpha\beta}$ for each
spin-isospin component
\begin{equation}
{\cal E}_{\alpha\beta} = \frac{1}{N} \sum_{k=1}^N | \psi_{v,\alpha\beta}
           ({\bf R}_k) - \psi_{F,\alpha\beta} ({\bf R}_k) |^2 \ ,
\label{eq:eab}
\end{equation}
where $N$ is the total number of configurations considered.
Finally, we define an average deviation ${\cal E}$
\begin{equation}
{\cal E} = \frac{1}{N} \sum_{k=1}^N {\cal E} _{ {\bf R}_k} \ .
\label{eq:e}
\end{equation}
The exact $\Psi_0({\bf R})$ satisfies $H \Psi_0({\bf R}) = E_0 \Psi_0({\bf R})$
at all ${\bf R}$, however the $\Psi_F({\bf R})$ does not satisfy this equation
in all regions of configuration space, presumably due to the truncation in
sums over partial waves and to interpolation between mesh points.
In some regions $\Psi_F^{\dagger}({\bf R}) H \Psi_F({\bf R})$ can be
different from $E_0\Psi_F^{\dagger}({\bf R})\Psi_F({\bf R})$ by as much as
25\%.
For the study of the deviations listed above, a set of 1,000 configurations
${\bf R}_k$ for which $\Psi_F^{\dagger}({\bf R}) H \Psi_F({\bf R})$ is within
5\% of $E_0 \Psi_F^{\dagger}({\bf R}) \Psi_F({\bf R})$ was chosen from a larger
set of random configurations distributed with the probability
$\Psi_F^{\dagger}({\bf R}) \Psi_F({\bf R})$.

We studied a wide variety of three-body correlations of the form
$\xi({\bf R}) O_{\xi}$.
For each $O_{\xi}$, values of $\xi({\bf R}_k)$ were obtained by minimizing
${\cal E}_{{\bf R}_k}$.  Simple functions of ${\bf R}$ were then chosen to
approximate the extracted values of $\xi({\bf R}_k)$.
In principle, we can choose a set of sixteen operators $O_{\xi}$ and define
the complete three-body correlation as:
\begin{equation}
     |\Psi_v \rangle = (1 + \sum_{\xi} \xi ({\bf R}) O_{\xi} ) |\Psi_p \rangle
\end{equation}
instead of using eq.\ (15).
The $\xi ({\bf R})$ can then be calculated by solving the matrix equation:
\begin{equation}
     \left( \sum_{\xi} \xi ({\bf R}) O_{\xi} \right)
     \psi_{p,\alpha\beta} ({\bf R}) =
     \psi_{F,\alpha\beta} ({\bf R}) - \psi_{p,\alpha\beta} ({\bf R})\ ,
\end{equation}
obtained from $\Psi_v = \Psi_F$, at each value of ${\bf R}$.
However, all our attempts led to unacceptable, rapidly fluctuating, functions
$\xi ({\bf R})$ presumably dominated by the small differences between $\Psi_F$
and the exact $\Psi_0$ and inappropriate choices of the sixteen operators
$O_{\xi}$.
In contrast, the $\xi ({\bf R})$ obtained by minimizing the deviation
${\cal E}_{{\bf R}_k}$ are smooth and useful for some of the operators
$O_{\xi}$.

{}From the many operators $O_{\xi}$ considered only a few produced a
significant
reduction of the deviations; we list below the successful ones.
A spatial three-body correlation,
\begin{equation}
     f^{c}_{ijk} = 1 + q^{c}_{1}({\bf r}_{ij}\cdot{\bf r}_{ik})
                                ({\bf r}_{ji}\cdot{\bf r}_{jk})
                                ({\bf r}_{ki}\cdot{\bf r}_{kj})
                                exp(-q^{c}_{2}R_{ijk}) \ ,
\label{eq:fcijk}
\end{equation}
reduces the probability for particles $i$, $j$, and $k$ to be in a line for
$q^c_1 > 0$.
The parameters of the new three-body correlations of type $x$ are denoted by
$q^{x}_{i}$; thus $q^{c}_{1}$ and $q^{c}_{2}$ are the parameters of
$f^{c}_{ijk}$.

A spin-orbit three-body correlation
\begin{equation}
 U^{\ell s}_{ijk} = \sum_{cyc} i\sigma_{i}\cdot({\bf r}_{ij}\times{\bf r}_{ik})
                    \left[ \bar v_{ik}(\bar f_{ij}'-\bar f_{jk}')
                         - \bar v_{ij}(\bar f_{ik}'-\bar f_{jk}')\right] \ ,
\label{eq:uls}
\end{equation}
where
\begin{eqnarray}
 &&  \bar f'(r) = q^{\ell s}_{1}exp(-q^{\ell s}_{2}r^{2}) \  \\
 &&  \bar v(r) = exp(-q^{\ell s}_{3}r^{2}) \
\end{eqnarray}
may be generated by the two-body spin-orbit interaction
$\bar v (r_{ij}) ({\bf L}\cdot{\bf S})_{ij}$ operating on the Jastrow
correlations $\bar f (r_{ik})$ and $\bar f (r_{jk})$.
It is expressed as a function of $\bar v$ and $\bar f$ to underline its
motivation, however $\bar v$ differs significantly from the bare
$\bf L \cdot \bf S$ potential and both $\bar v$ and $\bar f$ are determined
by minimizing the ${\cal E}_{{\bf R}_k}$.

An isospin three-body correlation
\begin{equation}
     U^{\tau}_{ijk} = \sum_{cyc} ({1 \over 3}R_{ijk} - r_{ij}) \left\{
                      q^{\tau}_{1}exp\left[-q^{\tau}_{2}
                      (X_{ijk} - q^{\tau}_{3})^{2}\right]
                    -2q^{\tau}_{1}exp(-2q^{\tau}_{2}X_{ijk}^{2}) \right\}
                      \tau_{i}\cdot\tau_{j} \ ,
\label{eq:utau}
\end{equation}
with
\begin{equation}
     X_{ijk} = \left[1+q^{\tau}_{4}
                       ({\bf \hat r}_{ij}\cdot{\bf \hat r}_{ik})
                       ({\bf \hat r}_{ji}\cdot{\bf \hat r}_{jk})
                       ({\bf \hat r}_{ki}\cdot{\bf \hat r}_{kj})\right]
               R_{ijk} \ .
\end{equation}
enhances the probability for $t_{ij} = 0$ when nucleons $i$ and $j$ are far
from $k$.

An example of a three-body correlation that we tried and found to be of
marginal utility is given by:
\begin{equation}
     U^{xt}_{ijk} = \sum_{cyc} i\sigma_{i}\cdot({\bf r}_{ij}\times{\bf r}_{ik})
                    {\bf \hat r}_{ij}\cdot{\bf \hat r}_{ik}
                    \left[ \bar g_{ij} \bar g_{ik} \right]
                    i\tau_{i}\cdot(\tau_{j}\times\tau_{k}) \ ,
\label{eq:uxt}
\end{equation}
where
\begin{equation}
     \bar g(r) = q^{xt}_{1}exp(-q^{xt}_{2}r^{2}) \ .
\end{equation}
Such a correlation may be generated by tensor interactions between pairs $ij$
and $ik$.
This correlation was omitted in our final energy calculations.

Finally, an improved parametrization of $f^p_{ijk}$ was found, and
eq.(\ref{eq:fpijk}) is replaced by:
\begin{equation}
     f^{p}_{ijk} = 1 - q^{p}_{1}(1-{\bf \hat r}_{ik}\cdot{\bf \hat r}_{jk})
                                exp(-q^{p}_{2}R_{ijk}) \ .
\label{eq:fpijkl}
\end{equation}
Initial values for the various parameters were obtained by minimizing
${\cal E}_{{\bf R}_k}$ for the test set of 1,000 Faddeev configurations.
The final values of the parameters, which are somewhat different, are given in
Table I; they were obtained by minimizing the binding energy of $^3$H with
the Argonne $v_{14}$ and Urbana model VIII interactions for samples of 10,000
variational configurations.

To illustrate the procedure used to determine the form of the new correlation
functions, we show in Fig.\ 1 the data points for the three-body central
correlation $f^{c}_{ijk}({\bf R})$ obtained by minimizing ${\cal E}_{{\bf
R}_k}$
at each value of ${\bf R}$.
These are for isosceles configurations $r_{13}=r_{23}$ with $R_{ijk}=6$ fm,
plotted against $cos\theta$, where $\theta$ is the angle formed by
${\bf r}_{13}$ and ${\bf r}_{23}$.
The functional form of $f^c_{ijk}$, eq.(\ref{eq:fcijk}), was chosen because it
can fit these numerical values with appropriate parameters $q^c_1$ and $q^c_2$.
The curve in Fig.\ 1 shows $f^{c}_{ijk}({\bf R})$ with the slightly different
values of $q^c_1$ and $q^c_2$ that minimize the energy; the data points and
curve are quite similar.

The improvement achieved in the main components of $\Psi_v$ can also be seen
in Fig.\ 2 where we plot the difference $\psi_{F,31}- \psi_{v,31}$ for
equilateral configurations as a function of $R_{ijk}$.
The $\psi_{F,31} \times 10^{-2}$ is also shown in Fig.\ 2 for comparison.
In this calculation the particles are placed in the xz-plane, where
$\psi_{\alpha\beta}$ are real, and for equilateral configurations
$\psi_{31}=-\psi_{12}$ holds exactly for antisymmetric $\Psi$.
The difference $\psi_{F,31}-\psi_{v,31}$ is $^{<}_{\sim}$ 1\% of $\psi_{F,31}$
for these configurations.

The average values of $|\psi_{F,\alpha\beta}|^2$, ${\cal E}_{\alpha\beta}$
[eq.(\ref{eq:eab})], $\Psi^{\dagger}_F \Psi_F$ and the total deviation
$\cal E$ [eq.(\ref{eq:e})] for the test set of 1,000 Faddeev configurations
are listed in Table II for the new and old $\Psi_v$.
They are normalized so that $\Psi^{\dagger}_F \Psi_F$ has unit average value
for these configurations; note that this does not correspond to the standard
normalization $\langle \Psi_F | \Psi_F \rangle = \langle \Psi_v | \Psi_v
\rangle
= 1$, however, it clearly exhibits the relative contributions and errors.

The main components, $\alpha\beta$ = 31 and 12, have a rather small relative
error, defined as ${\cal E}_{\alpha\beta}/
\overline{|\Psi_{F,\alpha\beta}|^2}$,
of $\sim 6\times 10^{-5}$ in the new $\Psi_v$.
Next in importance is the group of $s=\frac{3}{2}$ states ($\alpha$ = 5 to 8)
which have total orbital angular momentum $L_{tot} = 2$ and account for
$\sim 12\%$ of $\overline{\Psi^{\dagger}_F \Psi_F}$.
The largest of these have $\alpha\beta$ = 61 and 72, and a relative error of
$\sim 3\times 10^{-4}$.
The states $\alpha\beta$ = 71 and 51 have smaller contributions, and a larger
relative error of $\sim 10^{-2}$.
The next group has $s=\frac{1}{2}$, $m=\frac{1}{2}$ where the pair of nucleons
1 and 2 is in an odd parity state.
These states account for $\sim 0.1\%$ of $\overline{\Psi^{\dagger}_F \Psi_F}$
and have the largest relative error of $\sim 4\times 10^{-2}$.
The last group has $s=\frac{1}{2}$ and $m=-\frac{1}{2}$, for an $L_{tot} = 1$.
It also accounts for $\sim 0.1\%$ of $\overline{\Psi^{\dagger}_F \Psi_F}$, and
these components of the new $\Psi_v$ have an error of $\sim 2\times 10^{-2}$.

The ${\cal E}_{\alpha\beta}$ of the new $\Psi_v$ are smaller than those of
the old $\Psi_v$ for all values of $\alpha\beta$ other than 62.
The total ${\cal E}$ of 0.00062 is reduced to 0.00047 by the new $f^c_{ijk}$.
It is further reduced to 0.00035 by the inclusion of $U^{\ell s}_{ijk}$.
The subsequent reduction to ${\cal E}$ = 0.00028 is due to $U^{\tau}_{ijk}$.
The $U^{xt}_{ijk}$ correlations and the changes in the $f^p_{ijk}$ correlations
have very small ($< 10^{-5}$) effect on the total ${\cal E}$.

\section{BINDING ENERGY RESULTS FOR THE THREE- AND FOUR-BODY SYSTEMS;
CONCLUSIONS}

As mentioned in the previous section, the final values of the parameters of
the new three-body correlations were determined by minimizing the energy.
In Table III, we present the results for the binding energy of $^3$H, for the
Faddeev wave function and various variational wave functions moving from
the old to the new.
A set of 50,000 configurations sampled from $\langle \Psi_F|\Psi_F \rangle$
was used for these calculations.
The first column contains the values of the binding energy and the second the
difference $E_F - E_v$.
We can clearly see that this difference is reduced approximately by a factor
of two, when one goes from the old to the new $\Psi_v$, and that the
$f^{c}_{ijk}$ and $U^{\ell s}_{ijk}$ give almost 80\% of the improvement.

The last column of Table III gives $\langle\Delta | \Delta\rangle$, where
\begin{equation}
     |\Delta\rangle = |\Psi_F\rangle - |\Psi_v\rangle \ ,
\end{equation}
and $\langle\Psi_F | \Psi_F\rangle = \langle\Psi_v | \Psi_v\rangle = 1$.
The $|\Delta\rangle$ gives the admixture of excited states in $|\Psi_v\rangle$.
The mean energy of these excited states is given by:
\begin{equation}
     \overline{E_x} \sim \frac{E_v - E_0}{\langle\Delta | \Delta\rangle}
                    \sim 150 \ MeV
\end{equation}
for both old and new $\Psi_v$.
For this reason, the GFMC energy:
\begin{equation}
     [E(\tau)] = \frac{ \langle\Psi_v| H exp[-(H-E_0)\tau] |\Psi_v\rangle }
                      { \langle\Psi_v| exp[-(H-E_0)\tau] |\Psi_v\rangle } \ ,
\end{equation}
of $^3$H (and also $^4$He) converges to the true $E_0$ by a rather small
$\tau \sim 0.02$ MeV$^{-1}$.  In contrast the $\Psi_v$ of six-body nuclei
seem to have admixtures of states with much smaller $E_x$, and their $E(\tau)$
does not converge so rapidly to $E_0$ as $\tau$ increases~\cite{PPCW95}.

These new correlations, obtained by comparing the $\Psi_F$ and $\Psi_v$ for the
Argonne $v_{14}$ and Urbana VIII Hamiltonian, seem to have more general
applicability.
Results of variational calculations for $^3$H and $^4$He, with different
interaction models, using either the old $\Psi_v$ (Ref. [2]) or the new
$\Psi_v$ with the same three-body correlation parameters given in Table I
are listed in Tables IV and V.
(The difference between the old Argonne $v_{14}$ and Urbana VIII value
given in Table IV and the old $f^p_{ijk}$ given in Table III is of marginal
statistical significance; it can be attributed to slight changes to the
the $|\Psi_p\rangle$ in the present work compared to Ref. [2], and the
different random walks followed in the two calculations.)
A significant improvement is seen in each case, with $\sim$ 30--50\% of
the energy difference with exact calculations, where available, recovered.
In principle, the parameters of the new correlations should be optimized
separately for each case, but this has not been done in the present work.
The Faddeev energies shown in Table IV have been calculated from the Los
Alamos-Iowa $\Psi_F$ using Monte Carlo integration, and thus have a quoted
statistical error.
These energies are consistent with those published by the Los Alamos-Iowa
group~\cite{CPFG85}.
For $^4$He, the current VMC results are fairly competitive with other
recently developed methods, Faddeev-Yakubovsky (FY)~\cite{GK93} and
correlated hyperspherical harmonics (CHH)~\cite{VKR95}, as shown in Table V.
However, all these methods fall short of the exact GFMC calculations~\cite{C91}
in $^4$He.

If the wave function $|\Psi\rangle$ were an exact eigenstate of H, then it
should satisfy the eigenvalue equation:
\begin{equation}
     H |\Psi \rangle = E |\Psi \rangle
\label{eq:eigen}
\end{equation}
and this equation should be satisfied for any space configuration, meaning that
any local energy
\begin{equation}
     E ({\bf R}) = \frac {\Psi^{\dagger}({\bf R}) H \Psi({\bf R})}
                         {\Psi^{\dagger}({\bf R})\Psi({\bf R})}
\label{eq:eigenrr}
\end{equation}
should be equal to the eigenvalue $E$.
For the three-body system we considered sets of different space configurations,
each set being composed of configurations having the same $R_{ijk}$ (the
perimeter of the triangle formed by the three particles), and computed the
average of the energies returned by eq.(\ref{eq:eigenrr}).
These average values, $\bar E$, are then a function of $R_{ijk}$ and can be
plotted against it.
In Figs.\ 3, 4, and 5 we show the results respectively for the Faddeev wave
function, and the old and new variational wave function.
These calculations were done with the Argonne $v_{14}$ two-body and Urbana
model VIII three-body interactions.

{}From Fig.\ 3 we can clearly see that, even for the Faddeev wave function,
there
are regions, in particular for small and large perimeters, where the average
values differ from the energy expectation value.
Moreover, the variance is finite at all $R_{ijk}$, whereas it would be zero
for an exact eigenfunction.
The deviations of $\bar E (R_{ijk})$ and the variance of the Faddeev local
energy are perhaps due to the errors introduced by numerical interpolation
and the application of the permutation operators, required to go from the
Faddeev amplitude to the full wave function, as explained in the previous
section.
The application of these permutation operators in the Los Alamos-Iowa Faddeev
solution is implemented by means of partial wave series truncated at 34
channels.
This problem led us to consider, in the procedure to search for the three-body
correlations, only those regions in configuration space where the Faddeev wave
function gives an $E(R)$ close to the exact energy.
A comparison of the old and new variational wave function results (Fig.\ 4
to Fig.\ 5) shows that the new correlations clearly tend to push the average
values in the right direction and to reduce the variance.

In variational calculations of six-body nuclei~\cite{PPCW95} the new
three-body correlations continue to lower the variational upper bound.
One simple generalization that can be made is to let the $f^c_{ijk}$
correlation
have different strengths depending on whether none, one, or two of the
nucleons $i$, $j$, and $k$ are in the p-shell.
At present it appears best to let this correlation be zero if one or two are
in the p-shell, although an alternate structure might be even better in
these cases.
It is also beneficial to reduce the overall strength of the $U^{\ell s}_{ijk}$
and $U^{\tau}_{ijk}$ correlations.
With these steps, the new three-body correlations lower the energy by $\sim$
0.25 MeV in $^6$Li, and also reduce the variance.

In conclusion we have reported significant progress in the construction of
accurate variational wave functions for the nuclear few-body problem.
We believe that the new three-body correlations will lead to clear improvement
in GFMC wave functions for light nuclei, as well as in variational wave
functions of heavier systems.

\acknowledgements
We wish to thank J.L.\ Friar, B.F.\ Gibson, and G.L.\ Payne for the use of
their Faddeev wave functions.
One of the authors (AA) would like to thank the kind hospitality of the Physics
Department of the University of Illinois at Urbana-Champaign, where part of
this work has been performed.
The work of AA is supported by the University of Lisbon, Junta Nacional de
Investiga{\c c}\~ao Cient{\'{\i}}fica e Tecnol\'ogica under contract No.
PBIC/C/CEN/1108/92, that of VRP by the National Science Foundation via grant
PHY 94-21309, and that of RBW by the U.S.\ Department of Energy, Nuclear
Physics Division, under contract No.\ W-31-109-ENG-38.

\begin{figure}
\caption{The three-body central correlation, $(f^{c}_{ijk} - 1)$ for isosceles
configurations with $R_{ijk}=6fm$, as a function of $cos\theta$, where $\theta$
is the angle formed by $\vec r_{13}$ and $\vec r_{23}$.
The dots are the numerical values obtained by minimizing ${\cal E}_R$ and
the full line shows the analytical function with parameters that minimize
the energy.}
\label{fig1}
\end{figure}

\begin{figure}
\caption{The difference  $\psi_{F,31} - \psi_{v,31}$, for equilateral
configurations, as a function of $R_{ijk}$.
The dots and the diamonds correspond respectively to the old and new
variational wave functions.
The full line shows $\psi_{F,31} ({\bf R}_{ijk}) \times 10^{-2}$.}
\label{fig2}
\end{figure}

\begin{figure}
\caption{Local energy E(${\bf R}$) of $^3$H for 50,000 configurations of the
Faddeev wave function, sampled with probability $|\Psi_F|^2$ and binned
according to the perimeter variable $R_{ijk}$; the error bar denotes the
variance in each bin.
The horizontal line shows the average expectation value and the curve at
the bottom shows the relative distribution of the samples as a function of
$R_{ijk}$.}
\label{fig3}
\end{figure}

\begin{figure}
\caption{Local energy E(${\bf R}$) of $^3$H for 50,000 configurations of the
old variational wave function; notation as in Fig.\ 3.}
\label{fig4}
\end{figure}

\begin{figure}
\caption{Local energy E(${\bf R}$) of $^3$H for 50,000 configurations of the
new variational wave function; notation as in Fig.\ 3.}
\label{fig5}
\end{figure}

\narrowtext
\begin{table}
\caption{Values of parameters in three-body correlation functions.}
\begin{tabular}{cdl}
  $q^{c}_{1}$      &    0.19   & fm$^{-6}$         \\
  $q^{c}_{2}$      &    0.83   & fm$^{-1}$         \\
  $q^{p}_{1}$      &    0.20   &                   \\
  $q^{p}_{2}$      &    0.06   & fm$^{-1}$         \\
  $q^{\ell s}_{1}$ &  --0.12   & fm$^{-2}$         \\
  $q^{\ell s}_{2}$ &    0.12   & fm$^{-2}$         \\
  $q^{\ell s}_{3}$ &    0.85   & fm$^{-2}$         \\
  $q^{\tau}_{1}$   &  --0.014  & fm$^{-1}$         \\
  $q^{\tau}_{2}$   &    0.016  & fm$^{-2}$         \\
  $q^{\tau}_{3}$   &    1.2    & fm                \\
  $q^{\tau}_{4}$   &    0.25   &
\end{tabular}
\label{constants}
\end{table}

\mediumtext
\begin{table}
\caption{Average values of the wave functions and deviations.}
\begin{tabular}{lcccrccc}
$\alpha\beta$ & $s_{12}$ & $t_{12}$ & s & m & $\overline{|\Psi_{F,\alpha\beta}|
^2}$ & ${\cal E}_{\alpha\beta, new}$ & ${\cal E}_{\alpha\beta, old}$ \\
\tableline
31  & 1 & 0 & $\frac{1}{2}$ & $ \frac{1}{2}$ & $4.4\times 10^{-1}$
    & $2.7\times 10^{-5}$ & $1.2\times 10^{-4}$ \\
12  & 0 & 1 & $\frac{1}{2}$ & $ \frac{1}{2}$ & $4.4\times 10^{-1}$
    & $2.7\times 10^{-5}$ & $1.1\times 10^{-4}$ \\

61  & 1 & 0 & $\frac{3}{2}$ & $ \frac{1}{2}$ & $4.1\times 10^{-2}$
    & $1.6\times 10^{-5}$ & $1.8\times 10^{-5}$ \\
72  & 1 & 1 & $\frac{3}{2}$ & $-\frac{1}{2}$ & $3.0\times 10^{-2}$
    & $7.2\times 10^{-6}$ & $2.2\times 10^{-5}$ \\
81  & 1 & 0 & $\frac{3}{2}$ & $-\frac{3}{2}$ & $1.3\times 10^{-2}$
    & $3.8\times 10^{-5}$ & $4.6\times 10^{-5}$ \\
52  & 1 & 1 & $\frac{3}{2}$ & $ \frac{3}{2}$ & $1.2\times 10^{-2}$
    & $1.8\times 10^{-5}$ & $5.1\times 10^{-5}$ \\
62  & 1 & 1 & $\frac{3}{2}$ & $ \frac{1}{2}$ & $1.2\times 10^{-2}$
    & $4.7\times 10^{-5}$ & $4.2\times 10^{-5}$ \\
82  & 1 & 1 & $\frac{3}{2}$ & $-\frac{3}{2}$ & $4.3\times 10^{-3}$
    & $1.4\times 10^{-5}$ & $2.4\times 10^{-5}$ \\
71  & 1 & 0 & $\frac{3}{2}$ & $-\frac{1}{2}$ & $2.2\times 10^{-3}$
    & $1.5\times 10^{-5}$ & $2.2\times 10^{-5}$ \\
51  & 1 & 0 & $\frac{3}{2}$ & $ \frac{3}{2}$ & $2.0\times 10^{-3}$
    & $1.7\times 10^{-5}$ & $5.0\times 10^{-5}$ \\

11  & 0 & 0 & $\frac{1}{2}$ & $ \frac{1}{2}$ & $4.6\times 10^{-4}$
    & $1.7\times 10^{-5}$ & $2.9\times 10^{-5}$ \\
32  & 1 & 1 & $\frac{1}{2}$ & $ \frac{1}{2}$ & $4.6\times 10^{-4}$
    & $1.7\times 10^{-5}$ & $2.9\times 10^{-5}$ \\

42  & 1 & 1 & $\frac{1}{2}$ & $-\frac{1}{2}$ & $3.9\times 10^{-4}$
    & $5.6\times 10^{-6}$ & $8.7\times 10^{-6}$ \\
21  & 0 & 0 & $\frac{1}{2}$ & $-\frac{1}{2}$ & $3.8\times 10^{-4}$
    & $5.4\times 10^{-6}$ & $8.7\times 10^{-6}$ \\
41  & 1 & 0 & $\frac{1}{2}$ & $-\frac{1}{2}$ & $2.5\times 10^{-4}$
    & $7.4\times 10^{-6}$ & $2.5\times 10^{-5}$ \\
22  & 0 & 1 & $\frac{1}{2}$ & $-\frac{1}{2}$ & $2.1\times 10^{-4}$
    & $9.2\times 10^{-6}$ & $2.2\times 10^{-5}$ \\
\tableline
sum &  &  &  &  & 1.0
    & $2.8\times 10^{-4}$ & $6.2\times 10^{-4}$ \\
\end{tabular}
\label{deviations}
\end{table}

\mediumtext
\begin{table}
\caption{Comparison between binding energy values calculated with
 $|\Psi_F\rangle$ and the various $|\Psi_v\rangle$ described in the text.}
\begin{tabular}{lccc}
Wave function  &  $E$  & $E_v - E_F$  & $\langle\Delta|\Delta\rangle$ \\
\tableline
Faddeev & -8.479 (12) & \\
new $f^p_{ijk} + f^c_{ijk} + U^{\ell s}_{ijk} + U^{\tau}_{ijk}$
  & -8.354 (16) & .125 (19)  &  .00079 \\
new $f^p_{ijk}$ + $f^c_{ijk}+U^{\ell s}_{ijk}$
  & -8.346 (16) & .133 (19)  &  .00099 \\
new $f^p_{ijk}$ + $f^c_{ijk}$
  & -8.305 (15) & .175 (19)  &  .00111 \\
new $f^p_{ijk}$
  & -8.260 (20) & .219 (23)  &  .00139 \\
old $f^p_{ijk}$
  & -8.253 (19) & .227 (22)  &  .00146 \\
\end{tabular}
\label{energy differences}
\end{table}

\mediumtext
\begin{table}
\setdec 0.00
\caption{Binding energy in MeV for $^{3}$H with old and new correlations.}
\begin{tabular}{cccccc}
              &              &                & Argonne $v_{14}$ & Argonne
 $v_{14}$
 & Argonne $v_{14}$ \\
  Hamiltonian & Reid $v_8$   & Argonne $v_{14}$ & + Tucson       & + Urbana VII
 & + Urbana VIII  \\
\tableline
  old$^{\rm a}$ &\dec 7.31 (2) &\dec 7.45 (1) &\dec 8.80 (3)   &\dec 8.79 (1)
 &\dec 8.21 (2)   \\
  new           &\dec 7.44 (3) &\dec 7.53 (2) &\dec 9.05 (2)   &\dec 8.95 (1)
 &\dec 8.35 (1) \\
  Faddeev       &\dec 7.59 (2) &\dec 7.70 (1) &\dec 9.33 (2)   &\dec 9.05 (1)
 &\dec 8.49 (1)
\end{tabular}
\tablenotetext[1]{Ref.~\protect\cite{W91}}
\label{3h}
\end{table}

\mediumtext
\begin{table}
\setdec 0.00
\caption{Binding energy in MeV for $^{4}$He with old and new correlations.}
\begin{tabular}{cccccc}
              &              &                & Argonne $v_{14}$ & Argonne
 $v_{14}$
 & Argonne $v_{14}$ \\
  Hamiltonian & Reid $v_8$   & Argonne $v_{14}$ & + Tucson       & + Urbana VII
 & + Urbana VIII  \\
\tableline
  old$^{\rm a}$ &\dec 23.62(6)  &\dec 23.54(4) &\dec 30.64(9)   &\dec 30.51(4)
 &\dec 27.23(6)   \\
  new           &\dec 24.01(8)  &\dec 23.80(6) &\dec 31.69(9)   &\dec 30.79(5)
 &\dec 27.63(5)   \\
  FY$^{\rm b}$  &               &\dec 23.90    &                &
 &                \\
  CHH$^{\rm c}$ &               &\dec 23.93    &                &
 &\dec 27.48      \\
  GFMC$^{\rm d}$&\dec 24.55(13) &\dec 24.2(2)  &                &
 &\dec 28.3(2)    \\
\end{tabular}
\tablenotetext[1]{Ref.~\protect\cite{W91}}
\tablenotetext[2]{Ref.~\protect\cite{GK93}}
\tablenotetext[3]{Ref.~\protect\cite{VKR95}}
\tablenotetext[4]{Ref.~\protect\cite{C91}}
\label{4he}
\end{table}

\end{document}